\documentclass[11pt]{article}
\usepackage{epsfig}
\usepackage{graphics}
\usepackage{longtable}

\begin{document}

\centerline{\large{\bf Gamma Ray Bursts -- A radio perspective}}

\vspace*{0.5cm}

\centerline{{\bf Poonam Chandra}}

\begin{center}
{\small \noindent National Centre for Radio Astrophysics, Tata Institute of Fundamental Research, \\
Pune  University Campus, PO Box 3, Pune
411007, India}
\end{center}

\vskip 0.5cm

{\bf Abstract} 

Gamma-ray bursts (GRBs) are extremely energetic events at
 cosmological distances.  They provide  unique laboratory to
 investigate fundamental physical processes under extreme conditions. 
Due to extreme luminosities, GRBs are detectable at very high redshifts
and  potential tracers of 
cosmic star formation rate at early epoch.   
While the launch of {\it Swift} and {\it Fermi} has increased our understanding of GRBs tremendously, many new questions have
opened up.  Radio observations of GRBs uniquely probe the 
energetics and environments of
the explosion. However,  currently only 30\% of the bursts
 are detected in radio bands. Radio observations
with upcoming sensitive telescopes will potentially increase the 
sample size significantly, and allow one to follow the individual bursts for
a much longer duration and be able to answer some of the important issues
related to true calorimetry, reverse shock emission and
environments around the massive stars exploding as GRBs in the early Universe. 

\section{Introduction}

Gamma Ray Bursts (GRBs) are non-recurring bright flashes of $\gamma$-rays lasting from seconds to minutes.
As we currently understand, in the standard GRB  model a compact central engine is responsible for accelerating and collimating the ultra-relativistic
jet-like outflows. 
The isotropic energy release
 in prompt $\gamma$-rays ranges from $\sim 10^{48}$ to $\sim 10^{54}$~ergs, e.g. \cite{cenko11}.
 While the  prompt emission spectrum is mostly non-thermal,  presence of thermal or quasi-thermal components 
 have been suggested for a handful of bursts \cite{Kumar:2014upa}. 
 Since the initial discovery of GRBs \cite{grb1973} till the discovery of GRB afterglows at X-ray, optical and radio wavelengths 
three decades later \cite{costa97, van97, wijers97, frail97}, the 
origin of GRBs  remained elusive. 
The afterglow emission confirmed that GRBs are  cosmological in origin,
ruling out multiple theories proposed favouring Galactic origin of GRBs,
e.g. \cite{gm95}.

 In the {\it BATSE} burst population, the durations of GRBs  followed a bimodal distribution -- short GRBs
with duration less than $2$~s and long GRBs lasting for more than $2$~s
\cite{Kouveliotou:1993yx}. 
Long GRBs are predominantly found in star forming regions of late type 
 galaxies \cite{Fruchter:2006py}, whereas, short bursts are seen in a all 
kinds of galaxies \cite{Fong:2009bd}.  Based on these evidences, the current 
 understanding is
 that the  majority of long GRBs originate
  in the gravitational collapse of massive stars \cite{Woosley:2006fn}, whereas at least a fraction of
   short GRBs form as a result of the merger of compact object binaries (see
   Berger et al.  \cite{Berger:2013jza} for a detailed review). 
  
GRBs are detectable at very high redshifts. 
The highest  redshift GRB is GRB 090429B with
a photometric redshift of $z=9.4$ \cite{cucchiara11}. 
However, the farthest known  spectroscopically confirmed GRB is GRB 090423 at a redshift of $z=8.23$ \cite{tanvir09}, indicating star formation
must be taking place at such early epoch in the Universe \cite{chandra10}.  
At the  same time, some GRBs at lower redshifts have   revealed association
 with type Ib/c broad lined supernovae e.g. GRB 980425 associated with  
 SN 1998bw \cite{kulkarni98}.

  Since the launch of the {\it Swift} satellite in November 2004 \cite{gehrels04}, the field of GRB 
has undergone a major revolution. 
Burst Alert Telescope (BAT) \cite{barthelmy05} on-board {\it Swift} has been localizing $\sim100$ GRBs per year \cite{geh09}. 
X-ray Telescope (XRT, \cite{burrows05}) and Ultraviolet/Optical Telescope (UVOT, \cite{roming05}) on-board {\it Swift} slew towards the
 BAT localized position within minutes and provide uninterrupted detailed light curve at these bands.
Before the launch of the {\it Swift}, due to the lack of dedicated instruments at X-ray and optical bands the
afterglow coverage was sparse, which is no longer the case.
 {\it Swift}-XRT has revealed  that central engine is capable of 
injecting energy
into  the  forward  shock  at  late  times \cite{dai98,
zm02,liang07}.

GRBs are collimated events. An acromatic jet break seen in all frequencies,
is an undisputed signature of it. However, the jet breaks  are seen only in a few {\it Swift} bursts, e.g. GRB 090426 \cite{nicuesa11}, GRB
130603B \cite{fong14}, GRB 140903A  \cite{troja16}. Many of the bursts have not shown jet breaks. It could be because {\it Swift} is largely detecting fainter bursts with an 
average redshift of $>2$,  much larger than the
detected by previous instruments \cite{geh09}. The faintness of the
bursts makes it difficult to see
jet breaks.
 Some of the 
GRBs  have also revealed chromatic jet breaks, e.g. GRB 070125 \cite{chandra08}.

An additional issue is the narrow coverage
of the {\it Swift}-BAT in 15--150 keV range. 
Due to the narrow bandpass, the uncertainties associated in energetics are much larger since one needs to extrapolate to 1--10,000 keV bandpass
to estimate the $E_{\rm iso}$, which is a key parameter to evaluate the total released energy and other relations.
Due to this constraint, it has been possible to catch only a fraction of traditional
GRBs. 

The {\it Swift} 
drawback was overcome by the launch of {\it Fermi} in 2008, providing observation over a broad energy range of 
over seven decades in energy coverage (8 keV--300 GeV). Large Area Telescope (LAT, \cite{Atwood09}) on-board {\it Fermi} is an imaging gamma-ray detector
in 20 MeV--300 GeV range with a field of view of about 20\% of the sky and Gamma-ray Burst Monitor (GBM) \cite{Meegan09} on-board {\it Fermi }
works in 150 keV--30 MeV, and can detect GRBs  across the whole of the sky. The highest energy photon detected from a GRB puts   a stricter  lower  limit  on  the  outflow  Lorentz
factor.
{\it Fermi} has
 provided useful constraints on the initial Lorentz factor owing to its high energy coverage, e.g. short GRB 090510 \cite{ackermann10}. This is because to avoid 
pair production,  the  GRB  jet  must  be
moving toward the observer with ultra-relativistic speeds.
Some of the key observations by Fermi had been: i) the delayed onset of high energy emission for both long and short GRBs
 \cite{Abdo2009_nature,Abdo2009_080916c,Abdo2009_090902B}, ii) Long lasting LAT emission  \cite{Ackermann2013}, iii)
  very high Lorentz factors ($\sim 1000$) inferred for the detection of LAT high energy photons \cite{Abdo2009_nature}, iv) 
  significant detection of multiple emission components such as thermal component in several bright bursts
  \cite{Guiriec2011,Axelsson2012,Burgess2014a}, and v) power law \cite{Abdo2009_090902B} or spectral cut off at high energies
   \cite{Ackermann2011}, in addition to the traditional Band function \cite{band93}.

While the GRB field has advanced a lot after nearly 5 decades of extensive research since the first discovery, 
there are many open questions about prompt emission, content of the outflow, afterglow emission,
microphysics involved, and detectability of the afterglow emission etc.
Resolving them would enable us to understand GRBs in more detail and also use them  to probe the early universe
as they are detectable at very high redshifts. 
With the recent discoveries of gravitational waves (GW)\cite{abott16a,abott16b}, a new era of Gravitational Wave Astronomy has opened. GWs are ideal to probe short GRBs as they are the
most likely candidates of GW sources 
with earth based interferrometers. 

In this paper, we aim to understand the GRBs with a radio perspective. Here we focus on limited problems which can be answered with more
sensitive and extensive radio observations and modeling. By no means, this review is 
exhaustive in nature. In \S \ref{afterglow}, we review the radio afterglow in general and out current understanding.
In \S \ref{open}, we discuss some of the open issues in GRB radio afterglows.  The \S \ref{conclusions} lists the conclusion.

\section{Afterglow Physics - A radio perspective and some milestones}
\label{afterglow}

\begin{figure}
\begin{center}
\begin{tabular}{cc}
\resizebox{8.7cm}{!}{\includegraphics{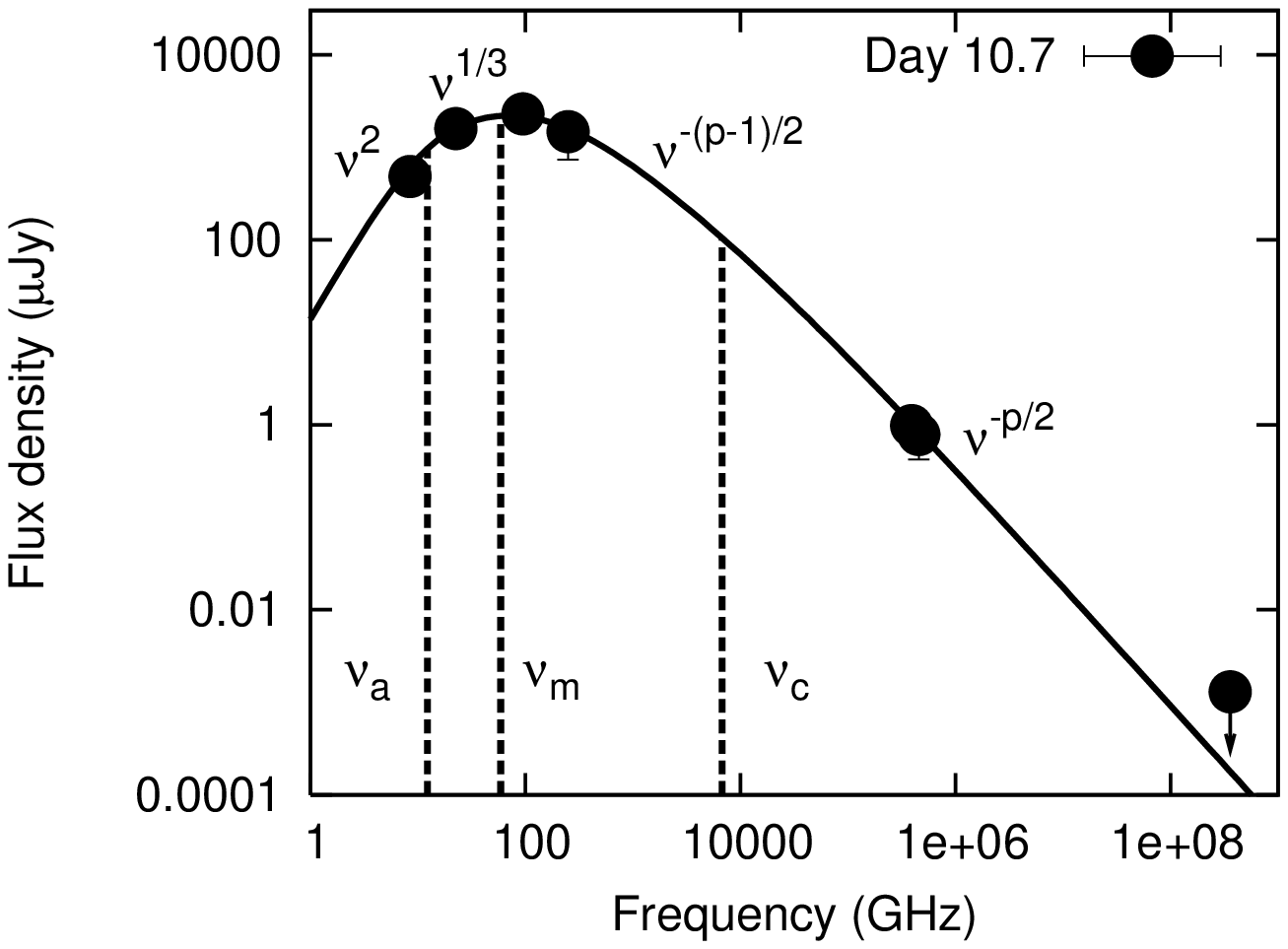}}&
\hspace*{-1.0cm} \resizebox{8.7cm}{!}{\includegraphics{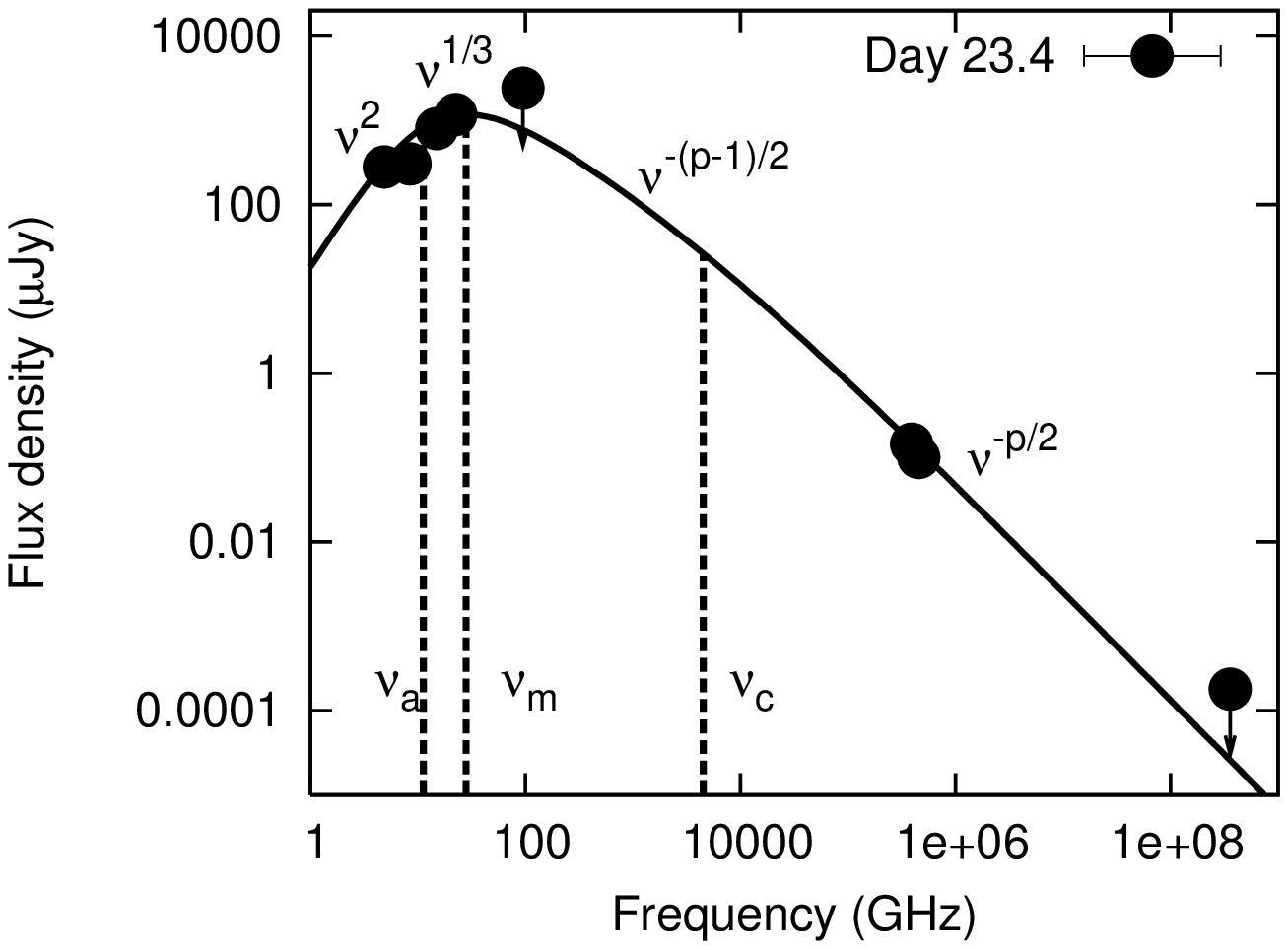}}\\
\end{tabular}
\caption{Multiwaveband spectra of GRB 070125 on day 10.7 and day 23.4. 
The spectra is in fast cooling regime. One can see that between the spectra on day 10.7 and 23.4, the peak has shifted to lower frequency.
The figure is reproduced from Chandra 
et al.\cite{chandra08}.
\label{070125}}
\end{center}
\end{figure}

In the standard afterglow emission model,  the relativistic ejecta interacting with the
circumburst medium gives rise to a forward shock moving into the ambient circumburst  medium and a reverse shock  going back into the ejecta.
The jet interaction with the circumburst medium   gives rise to mainly synchrotron emission in X-ray, optical and radio bands.  The peak of the spectrum 
moves 
from high to low observing frequencies over time due to the deceleration of the forward shock
 \cite{sari98}(e.g., see Fig. 
\ref{070125}). 
Because of the relativistic nature of the ejecta, 
the spectral peak is typically
below optical frequencies when the first observations commence,
  resulting in declining
light curves at optical and X-ray frequencies. However, optically rising light curves has been seen in a handful of bursts
after the launch of the {\it Swift} \cite{liang13}, e.g. GRB 060418 \cite{molinary07}.


The first radio afterglow was detected from GRB 970508 \cite{frail97}. Since then the radio studies of
GRB afterglows have increased our understanding of
the afterglows significantly, e.g.  \cite{chandra11,chandra12,granot14}. 
A major advantage of radio afterglow emission is that due to slow evolution, it peaks much later time and lasts
longer,  for  months  or  even  years (e.g., \cite{frail00, berger04, vander08}.
Thus unlike short lived optical or X-ray afterglows, radio observations present the possibility of following the full evolution of the fireball
emission from the very beginning till the non-relativistic phase (e.g., \cite{frail00, berger04, vander08}, also see GRB 030329 \cite{granot05, mesler13}. 
 Therefore, the radio regime plays
an important role in understanding the full broadband spectrum. This  constrains both the macrophysics
of the jet, i.e.  the energetics and the circumburst medium density, as well as the microphysics, such as energy imparted in electrons
and magnetic fields necessary for synchrotron emission \cite{wg99}.
Some of the phenomenon routinely addressed through radio observations are interstellar
scintillation, synchrotron self-absorption, forward shocks, reverse shocks, jet-breaks, non-relativistic transitions and obscured
star formation.

The inhomogenities in the local interstellar
medium manifest itself in the form of
interstellar scintillations and cause modulations in the radio flux density of a point source whose angular size is less than the characteristic angular size for scintillations \cite{goodman97}.   
GRBs are compact objects and one can see the  signatures of interstellar scintillation at early time radio observations, when  
 the angular size of the fireball is  smaller than the characteristic angular scale for interstellar scintillation. This 
 reflects in flux modulations seen in the radio observations. Eventually due to relativistic expansion, the
 fireball size exceeds the characteristic angular scale for scintillations and  the modulations quench. This can be utilised in
determining the source size and the expansion speed of the blast wave \cite{frail97}. 
 In GRB 970508 and GRB 070125, the initial radio flux density fluctuations were interpreted as interstellar 
scintillations, which lead to an estimation of the upper limit on the
fireball
size \cite{frail97, waxman98, chandra08}.  
In GRB 070125, the scintillation time scale and modulation intensity were consistent with those of diffractive scintillations, putting a
tighter constrains on the fireball size \cite{chandra08}.

Very Long Baseline Interferometry (VLBI) radio  observations  also play  a  key  role    by  providing
evidence for the relativistic expansion of the jet using  for bright GRBs.
This provides microarcsecond resolution and directly constrains the source size and its evolution.
So far  this has been possible for a nearby ($z=0.16$) GRB 030329 
\cite{taylor04}. In this case, the
 source size measurements were combined with its long term light
curves to better constrain the physical parameters \cite{granot05, mesler13}.
In addition, GRB 030329 also provided the first spectroscopic evidence for association of a GRB with a  supernova. 
This confirmed massive stars origin of at least a class of GRBs.

Radio observations are routinely used in broadband modelling of afterglows and used to derive blastwave parameters 
\cite{harrison01,pk01, yost03, chandra08, cenko11} (also see Fig. 
\ref{070125}).
Early radio emission is synchrotron self-absorbed, radio observations uniquely constrain the density of the circumburst medium. 
Radio studies have also proven useful for inferring the opening angles of the GRB jets as
their observational signature differs from those at higher wavelengths \cite{harrison99,
berger00, berger01, frail00}. 
Recently GRB 130427A, a  nearby,  high-luminosity event  was followed 
at all wavebands rigorously. It provided 
extremely good temporal (over 10 orders of magnitude) and spectral coverage, 
(16 orders
of magnitude in observing frequency \cite{ackerman14, maselli14}). 
Radio observations started as early as 8 hours \cite{laskar13}. 
One witnessed reverse shock and its peak  moving  from  high  to  low  radio  frequencies
over  time  \cite{laskar13, anderson14, perley14, vander14}.
 The burst is an ideal example to
 show   how early to late-time radio observations can contribute significantly to our
understanding of the physics of both the forward and reverse shocks.  

Radio afterglows can be detected   at high redshifts  \cite{frail06, chandra10} owing to the
negative
k-correction effect \cite{cl00}.  GRB 090423 at a redshift of 8.3 is the highest redshift (spectroscopically confirmed) known object in the Universe \cite{tanvir09}. It
was  detected in radio bands for several tens of days \cite{chandra10}.
The multiwaveband modeling indicated the $n~1$ cm$^{-3}$ 
density medium and the massive star origin of
the GRB. This suggested that the star formation was taking place even at a redshift of 8.3.

The radio afterglow, due to its long-lived nature, is able to probe the time when
the jet expansion has become subrelativistic and geometry has become  quasi-spherical \cite{frail00, frail05, vander08}, thus  can constrain energetics independent of geometry. 
This is possible only in radio bands as it lasts
  for  months  or  even  years (e.g. \cite{frail00, berger04, vander08}).
GRB 970508 remained bright more than a year after the discovery, when the ejecta had reached
sub-relativistic speeds.  This gave
the most accurate estimate of the kinetic energy of the burst \cite{frail00}.

Reverse shock probe the ejecta thus can poentially put constraints on
 the Lorentz factor and contents of the jet (e.g., \cite{perley14, anderson14}).
 The shock moving into the ejecta will result in an
optical flash in the first tens of seconds after the GRB under right conditions.
The radio regime is also well suited to probe the reverse shock emission as well. 
Short-lived radio flares, most likely due to
reverse shock, have also been detected from radio observations \cite{kulkarni99, berger03, nakar05, chandra10}
 and seems more common in radio bands than
in the optical bands. 
 GRB 990123 was the first GRB in which the reverse shock
was detected in optical \cite{akerlof99}
as well as in radio bands \cite{kulkarni99}.

From the radio perspective, GRB 030329 holds a very important place. 
It was the first high-luminosity burst at low redshift with a spectroscopic confirmation of a supernova associated with it. 
So far this is the only GRB
for which the source size has been measured with VLBI. The   radio afterglow of GRB 030329 was  bright and
long-lasting and  has been detected for almost a decade at  radio frequencies \cite{vander08, mesler12}.  
This enabled one to perform
broadband modeling in the different phases and has lead to tighter constraints on the physical parameters, \cite{granot05, mesler13}.
However, the 
absence of a counter jet poses serious
question in our understanding  of GRBs \cite{decolle12}.


\section{Open problems in GRB radio afterglows}
\label{open}

With various high sensitivity  new and refurbished telescopes e.g., Atacama Large Millimetre Array (ALMA), Karl J. Jansky Very Large
Array (JVLA), upgraded Giant Metrewave Radio Telescope (uGMRT) and upcoming telescopes e.g.,
Square Kilometre Array (SKA), the radio afterglow Physics of GRBs is entering into new era, where we can begin to answer some
of the open questions in the field, answers to which are long awaited. 
In this section, I discuss only some of those open problems in GRB science where radio measurements can play a crucial role.

\begin{figure}
    \centering
    \includegraphics[width=0.55\textwidth,natwidth=610,natheight=642]{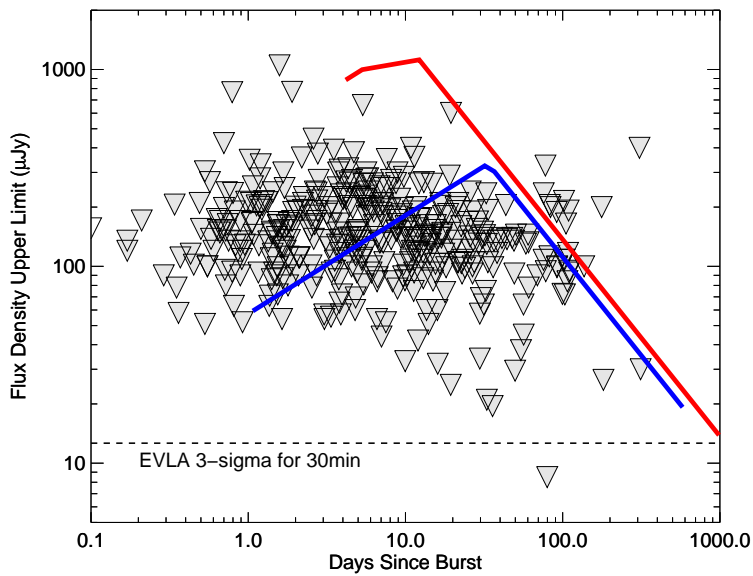}
\caption{Plot of $3-\sigma$ upper limits  at 8.5 GHz frequency band for all GRBs for which no
afterglow was detected. The red line shows light curve of a rare, bright event GRB 980703 and the blue line shows the light curve of a more typical event GRB 980329.
The detection fraction of radio afterglows in the first 10 days certainly appears
to be mainly limited by the sensitivity. The black dashed line indicates 
$3-\sigma$
sensitivity of the JVLA in its full capacity for a 30 minute integration time.
The figure is reproduced from \cite{chandra12}.}
\label{sensitivity}
\end{figure}

This review is not expected to be exhaustive. We concentrate on only a few major issues.

\subsection{Are GRBs intrinsically radio weak?}

Since the launch of the {\it Swift}, the fraction of X-ray and optically detected afterglows have increased
tremendously, i.e.  almost 93\% of GRBs have a detected X-ray afterglow \cite{evans09}, 
$\sim$ 75\% have detected optical afterglows  \cite{kann10,kann11}.
However, what is disconcerting is that the radio detection fraction have remained unchanged with only one-third of all
GRBs being detected in radio bands  \cite{chandra11,chandra12}. Chandra et al. \cite{chandra12} attributed it to sensitivity limitation of the current 
telescopes (see Fig. \ref{sensitivity}). This is because 
radio detected GRBs have flux densities typically ranging from a few tens of $\mu$Jy  to a few hundreds of $\mu$Jy \cite{chandra12}.
Even the largest radio telescopes have had the sensitivities close to a few tens of $\mu$Jy, making the radio afterglow detection 
sensitivity limited. 
The newer generation radio telescopes 
should  dramatically improve statistics of radio afterglows. For example, 
using numerical simulation of the forward shock, Burlon et al.
\cite{Burlon:2015yqa} predicts that the SKA-1 (SKA first phase) Mid band will be able to detect 
around $400 - 500$ 
radio afterglows per sr$^{-1}$ yr$^{-1}$. 

The
Five-hundred-meter Aperture Spherical radio Telescope (FAST)
\cite{zhang15,nan11,li13} is the largest worldwide single-dish radio telescope, being built in Guizhou province of  China with
an expected first light in Sep. 2016. FAST will continuously cover the radio frequencies between 70 MHz
and 3 GHz.  The radio afterglows of GRBs
is one of the main focus of FAST. 
Zhang et al. \cite{zhang15}
have estimated the detectability with FAST, of various GRBs
like failed GRBs, low luminosity GRBs, high luminosity GRBs and standard GRBs.
They predict that FAST will be able to detect most of the GRBs other than
subluminous ones upto a redshift of $z\le 10$.

However, Hancock et al. \cite{hancock13} used stacking of radio visibility 
data of many GRB  and their analysis still resulted in 
non-detection.  Based on this they proposed a class of GRBs which will
produce  intrinsically
faint radio afterglow emission and have black holes as their central engine. 
GRBs with magnetars as central engine will produce radio bright afterglow
emission. 
This is because 
the magnetar driven GRBs will have lower radiative efficiency and produce radio bright GRBs, whereas the black hole driven GRBs with their high radiative efficiency will use most
of their energy budget in prompt emission and 
will be radio-faint. This is a very 
important aspect and may need to be addressed. And if true, may reflect the nature of the central engine though radio measurements.
 JVLA at high 
 radio frequencies and the uGMRT at low radio frequencies test this hypothesis. 
SKA will eventually be the ultimate instrument to
distinguish between the sensitivity limitation versus the intrinsic dimness of radio bursts \cite{Burlon:2015yqa}.

\subsection{Hyperenergetic GRBs}

\begin{figure}
    \centering
    \includegraphics[width=0.55\textwidth,natwidth=610,natheight=642]{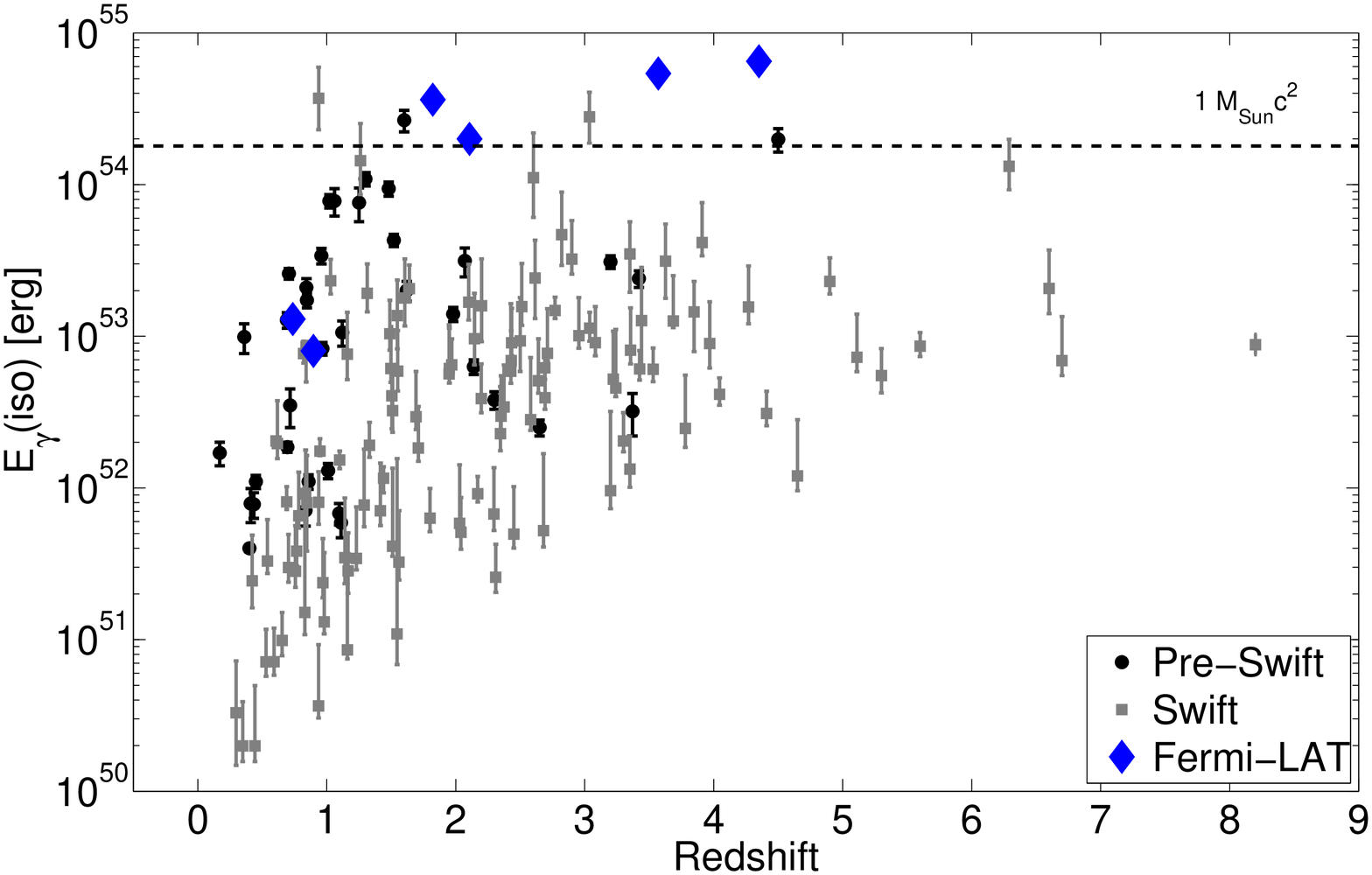}
\caption{Isotropic prompt gamma-ray energy release ($E_{\gamma,iso}$,
in rest-frame
1 keV--10 MeV bandpass),
of GRBs with measured redshift. One can see a large range of
of $E_{\gamma,iso}$. Reproduced from Cenko et al. \cite{cenko11}.
\label{cenko}
}
\end{figure}

Accurate  calorimetry  is  very important  to  understand  the true nature of the GRBs. 
This includes prompt radiation energy in the form of $\gamma$-rays
and kinetic energy in the form of shock powering the afterglow emission. 
Empirical constraints from models require that all long duration GRBs to have the
kinetic energies  $\le 10^{51}$ ergs. 
GRBs are collimated events, thus the jet opening angle is crucial to
measure the true budget of the energies.   While isotropic energies range of energies spread in four orders of magnitude (see Fig.
\ref{cenko}), the collimated nature of the jet makes the actual energies 
in much tighter range clustered around $10^{51}$ ergs \cite{frail01,berger03, bloom03}.
However,  it is becoming
increasingly evident that the clustering may not be as tight as envisaged 
and the actual energy range may be much wider than anticipated earlier. A population
of nearby  GRBs have relativistic energy orders of magnitude smaller
than a typical cosmological GRB; these are called subluminous GRBs, e.g. GRB 980425 \cite{soderberg04, liang07}.
 {\it Fermi} has provided evidence  for a class
of hyperenergetic GRBs. These GRBs have   total prompt  and kinetic energy release, inferred
via broadband modeling  \cite{pk02, yost03}, to be  at least an order of magnitude above the canonical value of $10^{51}$
erg \cite{chandra08, chandra12, cenko10, cenko11}.
The total energy budget of these hyperenergetic GRBs poses a significant challenge for some accepted progenitor
models. 
The
maximum energy release in magnetar models \cite{usov92} is $3\times10^{52}$ 
erg,    set  by  the  rotational  energy  of  a
maximally rotating stable neutron star \cite{thompson04,metzger07}. 

It has been very difficult to 
constrain the true prompt energy budget of the GRBs, mainly, for the
following reasons. So far, Swift has been instrumental in detecting 
majority of the GRBs. However, peak of the emission for various GRBs 
lie outside the narrow
energy coverage of Swift-BAT (15--150 keV).
In addition, extrapolation of 15--150 keV to 1--10,000 keV bandpass cause
big uncertainties in the determination of prompt isotropic energies. 
With its huge energy coverage (8 keV-300 GeV),
{\it Fermi} has overcome some of these limitations and
provided unparalleled constraints
on the spectral properties of the prompt emission.
{\it Fermi} has been able to distinguish the true hyperenergetic
bursts (such as GRB 090323, 090902B, 090926A, \cite{cenko11}, also
see Fig. \ref{cenko}).
While {\it Swift} sample is biased towards faint bursts, {\it Fermi} sample is
biased towards GRBs with very large isotropic energy releases ($10^{54}$
erg), which even after collimation correction reach very high energies e.g. \cite{cenko11,cenko06}, and
 provide some
of the strongest constraints on possible progenitor models. 

The uncertainty in jet structure in GRBs pose additional difficulty in constraining the energy budget of GRBs.
Even after a jet break is seen, to convert it into opening angle, one needs density 
to convert it into the collimation angle. While some optical light curves 
can be used to constrain the circumburst density (e.g. Liang et al. 
cite{liang13}), radio SSA peak is easier to detect due to slow evolution
in radio bands.  
With only one third of sample being radio bright, this has been possible for only a handful of bursts. A  larger radio sample at lower frequencies, 
at early times when synchrotron self absorption (SSA) is still playing a major role, 
could be very useful. The uGMRT after upgrade will be able to probe this regime as SSA will be affecting the radio emission at longer wavelength for
a longer time. 
However, the this works on the assumption that the entire relativistic outflow is collimated
into a single uniform jet.
While the  proposed
double-jet models for GRB 030329 \cite{berger03b,vander05} and GRB
080319B \cite{racusin08} ease out the  extreme efficiency
requirements, it  has caused additional concerns.  
   
 The ALMA also has an important role to play since GRB spectrum at early times
peak at mm wavelengths, when it is the brightest.  ALMA with its high sensitivity can detect such
events at early times and give better estimation of the kinetic energy of the burst.

While X-ray and optical afterglows stay above detection 
limits only for weeks or months, radio afterglows of nearby bursts can be detected upto 
years \cite{frail00,resmi05}.
The longevity of radio afterglows also make them interesting laboratories to study the 
dynamics and evolution of relativistic shocks. At late stages, the fireball would have expanded sideways
so much that it would essentially made transition into
non-relativistic regime and become quasi-spherical and independent of the jet geometry, calorimetry can be employed to 
  obtain the burst energetics
  \cite{frail00,vander08}.
These estimates will be free of relativistic effects and collimation corrections.
This regime is largely unexplored due to limited number  of bursts
staying above detection limit beyond sub-relativistic regime. 
Several numerical calculations exist for the afterglow evolution starting from the 
relativistic phase and ending in the deep non-relativistic phase \cite{vanEerten:2011bf, decolle12}.
 SKA with its $\mu$Jy  level sensitivity will be able to extend the current limits of afterglow longevity.
  This will provide us with an unprecedented opportunity to study the  non-relativistic regime of afterglow 
  dynamics and thereby will be able to refine our understanding of relativistic to non-relativistic transition of the blastwave 
  and changing shock microphysics and calorimetry in the GRBs. 
Burlon et al.  \cite{Burlon:2015yqa} have computed that SKA1-MID will be able
to observe  2\% afterglows till the non-relativistic (NR) transition. But that the full SKA will routinely observe 15\% of the whole
GRB afterglow population at the NR transition.

\begin{figure}
    \centering
    \includegraphics[width=0.41\textwidth,natwidth=610,natheight=642]{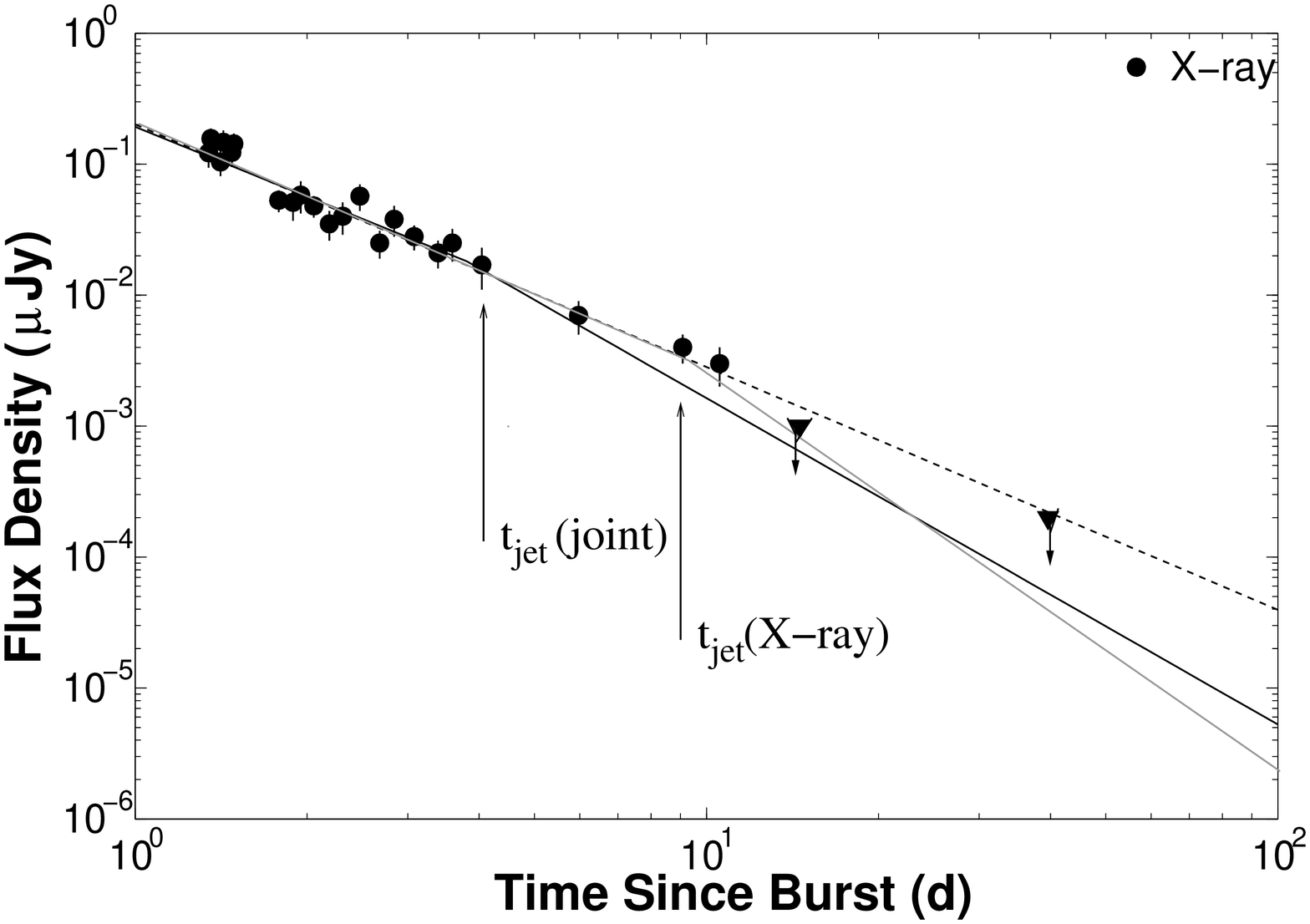}
    \includegraphics[width=0.53\textwidth,natwidth=610,natheight=642]{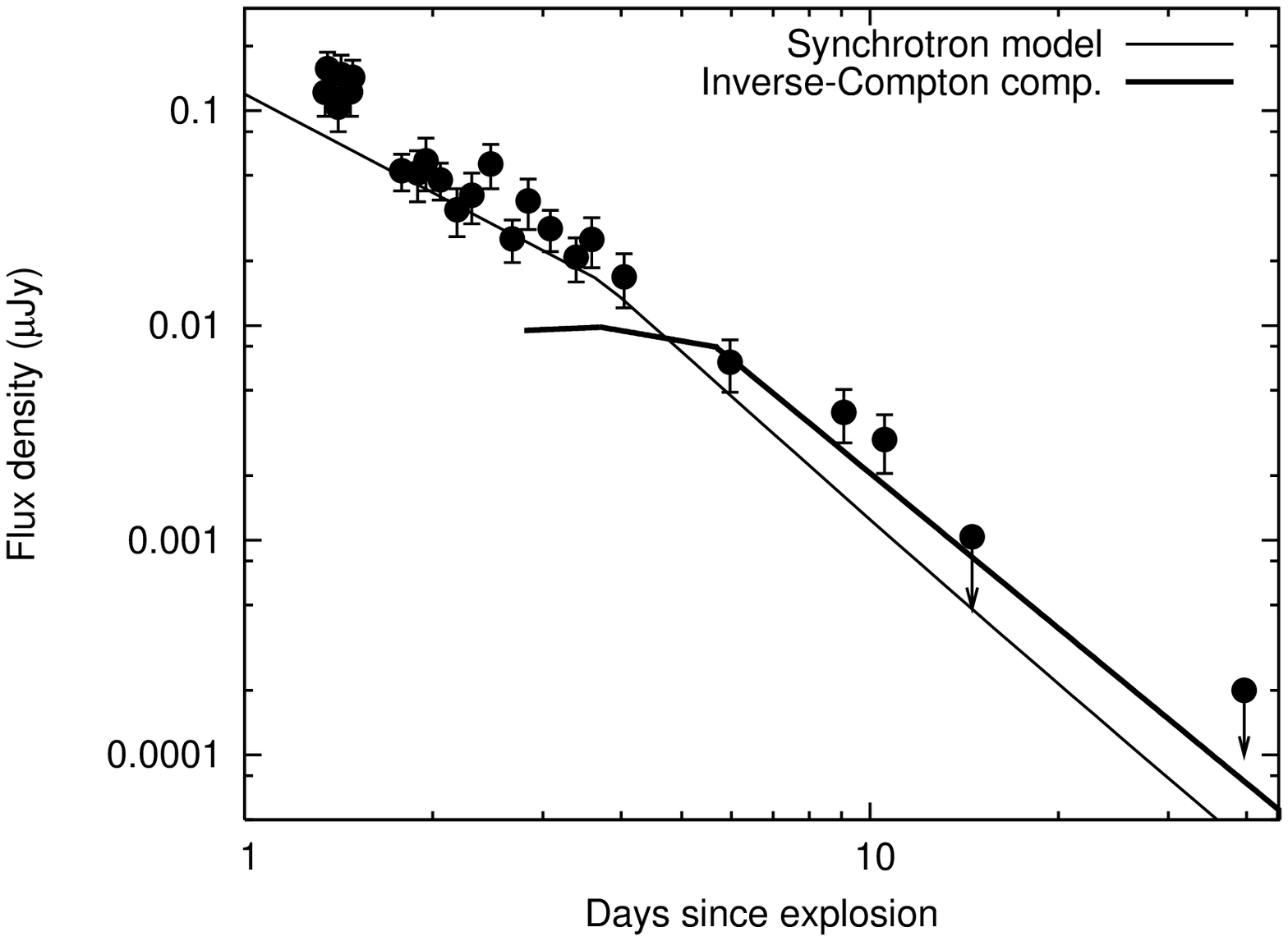}
\caption{\textit{Left}: X-ray light curve of GRB 070125. 
Best-fit single power-law models are shown with dashed lines, while the broken
power-law models are shown in solid lines.
The $t_{\rm jet}$(joint) is the
joint fit to optical and X-ray data and grey solid line
$t_{\rm jet}$(X-ray) is the independent
The independent fit
shifts the jet break to $\sim$ $9-10$ days, which was found to be day 3 for 
optical bands.
\textit{Right}: Contribution of IC in the synchrotron model for the X-ray
light curve of GRB 070125. The thin line represents
the broadband model with the synchrotron component only. The thick line represents
the IC light curve. One can see that IC effect can delay the
jet breaks in X-ray bands
 \cite{chandra08}.
\label{jet}
}
\end{figure}

\subsection{Can jet breaks be chromatic?}

After the launch of {\it Swift}, one obtained a far better sampled
optical and X-ray light curves, thus expected to witness achromatic jet breaks
across the electromagnetic spectrum, a robust signature
associated with a collimated outflow.
  Several groups
conducted a comprehensive analysis of a large sample of light curves 
of {\it Swift} bursts in the
X-rays
\cite{panaitescu07,kb08,racusin09,liang08}.
and optical \cite{wang15} bands. Surprisingly fewer Swift bursts have shown this unambiguous signature of
the jet collimation. Without these collimation
angle,  the true energy release from Swift events has remained highly uncertain.  A 
natural explanation for absence of the
jet breaks  can be attributed to the high sensitivity of
{\it Swift}. Due to its high sensitivity {\it Swift} is preferentially selecting GRBs with smaller isotropic gamma-ray energies and larger redshifts. 
This dictates that typical {\it Swift} events will  have large opening
angles, thus causing
jet breaks to occur at much time than those of pre-{\it Swift} events.
Since afterglow is already weak at later times,  making 
jet-break measurements is quite difficult
\cite{perna03,kb08}.

There have been some cases where chromatic jet breaks are also seen.  
For example, in GRB 070125, the X-ray jet break occurred around day 10, whereas the
optical jet break occured on day 3. Chandra et al. \cite{chandra08} attributed it to  inverse Compton
(IC) effect, which does not effect the photons at low energies, but shifts the X-ray jet break at a later
time (see Fig. \ref{jet}, \cite{chandra08}).  As IC effects are dominant in high density medium, radio observations are an important indicator of the effectiveness of
the IC effect.   \cite{chandra08} showed that for a given density of GRB
070125, the estimated delay in X-ray jet break due to the IC effect is consistent with the observed
delay.  However, this area needs to be explored further for other GRBs.
While high density bursts are likely to be brighter in radio bands, it may cause a burst to be a dark one in optical wavelength (Xin et al. 
\cite{xin10} and reference in there), which then make it difficult to detect the jet break simultaneously in several wavelengths. 
uGMRT and
JVLA will be ideal instruments to probe IC effect, and will potentially be
able to explain the cause of chromaticity in some of the {\it Swift} bursts.


\subsection{High-z GRBs and PoP III stars}

One of the major challenges of the
observational cosmology  is to understand the reionization of the Universe, when the first luminous sources were formed. 
So far quasar studies of the Gunn-Peterson absorption trough, the luminosity evolution of Lyman
galaxies, and the polarization isotropy of the cosmic microwave background
have  been used as diagnostics. But they have  revealed a complicated picture in which reionization took place over a range of redshifts.

 The ultraviolet emission from young, massive stars (see 
Fan et al. \cite{fan06} and references therein) appears to be the dominant source of reionization.
However,  none of these 
massive stars have been detected so far, 
Long GRBs, which are explosions of massive stars, are detectable out to large distances due to their extreme 
luminosities, and thus are the potential signposts of the 
early massive stars.  
GRBs are predicted to occur at redshifts beyond those where quasars are expected, thus they could be
used to study both the reionization history and the metal enrichment of the early Universe \cite{totani06}.
They could potentially reveal the stars that form from the first dark matter
halos through the epoch of reionization  \cite{lr00, cl00, gou04}.
The radio, infrared, and X-ray afterglow emission
from GRBs are in principle observable out to z=30 \cite{miralda98, lr00, cl00, gou04, im05}.
 Thus GRB afterglows make
ideal sources  to probe the intergalactic medium as well as the interstellar medium in their host
galaxies at high $z$. 

\begin{figure}
    \centering
    \includegraphics[width=0.40\textwidth,natwidth=610,natheight=642]{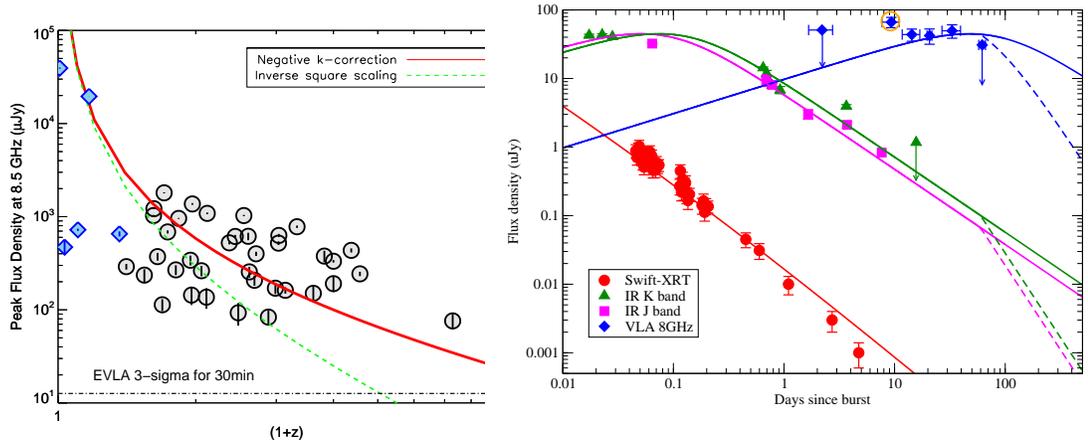}
    \includegraphics[width=0.42\textwidth,natwidth=610,natheight=642]{f5b.eps}
\caption{\textit{Left}: 
The 8.5 GHz radio peak flux density versus ($1+z$) plot for radio afterglows
  with known redshifts.    Blue diamonds are GRBs associated with supernovae, while 
  the gray circles denote cosmological GRBs.  
  The green dashed line indicates if the flux density scales as 
  the inverse square of the luminosity distance. The red thick line is
  the flux density scaling in the canonical afterglow model which
  includes a negative-$k$ correction effect, offsetting the diminution
  in distance (reproduced from \cite{chandra12}).
\textit{Right}: Multiwaveband afterglow modeling of highest redshift GRB 
090423 at $z=8.23$ (reproduced from
 \cite{chandra10}).
\label{redshift}
}
\end{figure}

 The fraction of detectable GRBs that lie at high redshift ($z>6$) is, however, expected to be less than 10\% \cite{perley09, bl06}.
So far there are only 3 GRBs with confirmed measured redshifts higher than 6. These are GRB 050904 \cite{kawai06}, 
GRB 080913 \cite{greiner09} and GRB 090423 \cite{tanvir09}.
 Radio bands 
are ideal to probe GRB circumburst environments at
high redshift because radio flux density show only a
weak dependence on the redshift, due to the negative
k-correction effect \cite{cl00} (also see \cite{chandra11} and Fig. \ref{redshift}).
In  K-corection  effect,  the  afterglow  flux  density  remains  high  because  of  the  dual  effects  of  spectral
and temporal redshift, offsetting the dimming due to the increase in distance \cite{lr00} (see Fig. \ref{redshift}. 
GRB 050904 and GRB 090423 were detected in radio bands
and radio observations of these bursts allowed us to put constraints on the density
of the GRB environments at such high redshifts.  While the density of GRB 090423 was $n\sim 1$ cm$^{-3}$ \cite{chandra10} (Fig. \ref{redshift}), the density of
GRB 050904 was $\sim 100$ cm$^{-3}$, indicating dense molecular cloud surrounding
the GRB 050904 \cite{gou07}. This revealed that  these two high-$z$ GRBs exploded in a very different environments.

ALMA  will be a potential tool for selecting
potential high-$z$ bursts that would be suitable for intense follow-up across the electromagnetic spectrum.  With an order of
magnitude enhanced sensitivity the JLAwill be able  to study a high-z
GRB for a longer timescale. For example, VLA can detect GRB 090423-like burst for almost 2 years. The uGMRT can also detect bright bursts upto a redshift of 
$z \sim 9$. These measurements will therefore obtain better density measurements,  and reveal the
environments where massive stars were forming in the early universe.

\subsection{Reverse shock}

 In a GRB explosion, there is
 a forward shock moving forward into the circumburst medium, as well as a reverse shock moving backwards
into the ejecta \cite{sari99}.  
The nearly self-similar behavior of a forward shock
means that little information is preserved about the central engine
properties that gave rise to the GRB. In contrast, the brightness of
the short-lived reverse shock depends  on the initial Lorentz
factor  and the magnetization of the ejecta.
Thus, multi-frequency observations of reverse shocks  tell  about the
acceleration, the composition and the strength and orientation of any
magnetic fields in the relativistic outflows from GRBs
\cite{kob00,zkm03,np04,perley14,anderson14}. 
In general, the reverse shock is expected to  result in an
optical flash in the first tens of seconds after the bursts \cite{akerlof99}.  difficult to detect it as  robotic telescopes are required for fast trigger. 

The discovery of a bright optical flash from GRB\,990123
\cite{akerlof99} lead to extensive searches for reverse shocks
\cite{akerlof00,rmf+06,raa+09,gkm+09} in optical bands. One expected to see more
evidences of reverse shocks in optical bands due to
{\it Swift}-UVOOT, however, based on these efforts it seems that
the
incidence of optical reserve shocks is low.
Since the peak of this emission
moves to lower frequencies over time and can be probed at radio frequencies on a time-scale of
hours to days \cite{kulkarni99}, the radio regime is  well suited
for  studying  early-time  reverse shock phenomena.

There have been several observational and theoretical studies of 
radio reverse shocks after the first reverse shock
detection in GRB 990123 \cite{kulkarni99}. Gao et al.
\cite{Gao:2013mia}, Kopac et al. \cite{Kopac:2015dia} and 
Resmi \& Zhang \cite{Resmi:2016a} have carried out comprehensive analytical and numerical calculations of radio 
reverse shock emission and about their detectability. 
It has been shown  \cite{chandra12, laskar13} that deep and fast monitoring campaigns of radio reverse shock emission could be achieved with the 
VLA for a number of bursts. JVLA radio frequencies are
well suited as  reverse shock emission is brighter in higher radio frequencies where self-absorption effects are relatively lesser. 
Radio afterglow
  monitoring campaigns in higher SKA bands (e.g. SKA1-Mid Band-4 and Band-5) will definitely be useful in exploring reverse shock characteristics
  \cite{Burlon:2015yqa}.

Reverse shock is detectable in high redshift GRBs ($z \ge6$) as well. 
Inoue et al. (2007) have predicted that at mm bands, the 
effects of time dilation almost compensate for frequency redshift.
Thus resulting in a near-constant
observed peak frequency  at a few hours post-event, and a flux
density at this frequency that is almost independent of redshift.
 Thus ALMA mm band is ideal to look for reverse shock signatures at high redshifts.   
 Burlon et al. \cite{Burlon:2015yqa} predict that 
 SKA1-Mid will be able to detect  a reverse shock from a  GRB990123  like GRB at a redshift of $\sim 10$.

\subsection{Connecting prompt to afterglow Physics}

 {\it Swift}  is an ideal instrument for quick 
  localization of GRBs and rapid follow up and 
consequently redshift measurement \cite{geh09, geh12}, and {\it Fermi} for the wide band spectral measurement during the prompt emission. 
However, good spectral and timing measurement covering  early prompt 
to late afterglow phase is available for a few sources and rarely 
available for the short GRBs.
Some of the key problems that can be addressed by the observation of the radio afterglows 
 in connection with the prompt emission is:
i) Comparing the Lorentz factor estimation with both LAT detected GeV photons as well as from the 
 reverse shock \cite{Kidd2014, Iyyani2013}. 
ii) Comparison between non-thermal emission of both the prompt as well as 
afterglow emission. This would enable one to constrain the microphysics of the shocks accelerating electrons to 
ultra-relativistic energies eventually producing the observed radiation.
iii) Detailed modelling of the GRB afterglows. This will enhance our knowledge
 about the circumburst medium surrounding the progenitors.
iv) Current refurbished and upcoming radio telescopes  with its finer sensitivity would play a key role in constraining the 
energetics of GRBs which is crucial in estimating the radiation efficiency of the prompt emission of GRBs. 
This would  strengthen the understanding of the hardness - intensity correlation 
\cite{Amati2002}.

The recently launched    {\it AstroSAT}  satellite \cite{singh14} carries several instruments  enabling
multi-wavelength studies.  The Cadmium Zinc  Telluride  Imager (CZTI) 
on-board {\it AstroSAT} can
provide time resolved polarisation measurements for bright GRBs and
can act as a monitor above  80 keV \cite{rao15, bhalerao15}.   So far no other 
instrument has such capability to detect polarization. 
 Hence, for a few selected bright GRBs, CZTI, in 
conjunction with  ground based observatories like  uGMRT and JVLA, and other space based facilities
can provide a complete observational
picture of a few bright GRBs from early prompt phase to late afterglow. This will provide us with
a comprehensive picture of GRBs.

\subsection{Some other unresolved issues}

So far I have discussed only that small fraction
of on-axis GRBs, in which  the jet is oriented along our line of sight. 
Due to large Lorentz factors, small opening angles of the collimated jets,  we only detect a small
fraction of of GRBs \cite{rhoads97}. 
Ghirlanda et al. \cite{ghirlanda14}  
 have estimated that for every GRB detected, there must be 260 GRBs
which one is not able to detect. 
However, their
existence can be witnessed as ``orphan afterglow'' at late times when the GRB jet is decelerated and spread laterally to come into our
line of sight.
 At such late times, the
emission is expected to come only in radio bands. 
So far attempts to find such orphan radio afterglows have been unsuccessful \cite{berger03, soderberg06, bietenholz14}.
Even if detected, disentangling the orphan afterglow emission from other classes
will be very challenging. Soderberg et al. \cite{soderberg06} carried out a survey towards the direction of 68 Type Ib/c supernovae
looking for the  orphan afterglows and put limit on GRB opening angles, $\theta_j>0.8$ d.
The detection of  population of orphan afterglows with upcoming sensitive radio facilities is promising. 
 This will give a very good  handle on jet opening angles, and on the 
 total GRB
rate whether beamed towards us or not.

The inspiral and merger of binary systems with black holes or neutron stars have been speculated as primary source of gravitational waves (GWs)
for the ground based GW interferrometers \cite{thorne87,schutz89}.
The discovery of GWs from GW 150914 \cite{abott16a} and GW 151226 \cite{abott16b} with the Advanced LIGO 
detectors have provided the first observational evidence of the binary black hole systems inspiraling and merging.  
At least some of the compact binaries involving a neutron star are expected to give rise
to radio afterglows of short GRBs.  Electromagnetic counterparts, including radio emission from a GW source 
is extremely awaited for it will for the first time confirm this hypothesis.
If localized at high energies, targetted radio observations can be
carried out to study these events at late epochs. 

Short GRBs arising from mergers of two neutron stars eject significant amount of mass in several components, including a 
sub-relativistic dynamical ejecta, mildly-relativistic shock-breakout, 
and a relativistic jet  \cite{hp15}. 
Hotokezaka \& Piran \cite{hp15} have calculated the expected radio signals produced
 between the different components of
the ejecta and the surrounding medium. The nature of radio emission years after GRB will provide invaluable information on the merger process \cite{hp15}
 and the central products \cite{fong16}.  Fong et al. \cite{fong16} have predicted that the formation of stable magnetar of energy
 $10^{53}$ erg during merger process will give rise to a radio transient a year later. They carried out  search for radio emission from 9 short 
 GRBs in rest frame times  of $1-8$ years and  concluded that such a magnetar formation can be ruled out in at least half their sample.

 In addition, radio observations can also probe the star formation and the metallicity of the GRB host galaxies when optical emissions are obscured by dust \cite{graham15,greiner16}.

\section{Conclusions}\label{conclusions}

In this article, I have reviewed the current status of the 
{\it Swift}/{\it Fermi} GRBs in context of their radio emission. With
 improved sensitivity of the
refurbished radio telescopes, such as JVLA and uGMRT and 
upcoming telescopes like SKA, it will be possible to answer many open
 questions. 
The most crucial of them is the accurate calorimetry of the GRBs.  
Even after observing a jet break in the GRB afterglow light curves, 
which is an unambiguous signature of the jet collimation, one needs density 
estimation to convert the jet break epoch to collimation angle. 
The density information
can be more effectively provided by the early radio measurements when 
the GRBs are still synchrotron self-absorbed. 
So far it has been possible for very limited cases
because only one third of the total 
GRBs have been detected in radio bands \cite{chandra12}.  Sensitive radio 
measurements are needed to understand whether the low 
detection rate of radio afterglows is intrinsic to GRBs or the sensitivity 
limitations of the current 
telescopes are playing a major role. In the era of JVLA, uGMRT, 
ALMA and upcoming SKA, this issue should be resolved.  In addition, 
these 
sensitive radio telescopes will be crucial to detect radio afterglows 
at very high redshifts and provide unique constraints on the 
environments of the
exploding massive stars in the early Universe.
 If GRBs are not intrinsically dim in radio bands and the sample is indeed sensitivity limited,  then SKA is expected to detect almost 100\% GRBs \cite{Burlon:2015yqa}. 
 SKA will be able to study the individual bursts in great detail. 
 This will also allow us to carry out various statistical analysis of the
 radio sample and drastically  increase our overall 
understanding of  the afterglow evolution from very early time to 
non-relativistic regime. Detection of the orphan 
afterglow is due any time and will be novel in itself. 

\vspace*{0.5cm}

\section{Acknowledgements}
I thank L. Resmi, Shabnam Iyyanni, A. R. Rao, Kuntal  Misra and D. Frail for many useful discussions in the past, which helped shape 
this article.
I acknowledge support from the Department of Science and Technology via SwaranaJayanti Fellowship award (file no.DST/SJF/PSA-01/2014-15).
I  acknowledge SKA Italy handbook (http://pos.sissa.it/cgi-bin/reader/conf.cgi?confid=215), where many of the SKA numbers on
sensitivity, GRB detection rates etc. are taken.

\end{document}